\relax
\documentclass[letterpaper]{article} 
\usepackage{aaai19}  
\usepackage{times}  
\usepackage{helvet}  
\usepackage{courier}  
\usepackage{url}  
\usepackage{graphicx}  
\begin{document}
%
\title{To Kavanaugh or Not to Kavanaugh: That is the Polarizing Question}
\author{Kareem Darwish\\
Qatar Computing Research Institute, HBKU\\
Doha, Qatar\\
KDarwish@HBKU.edu.qa\\
}
\maketitle
\begin{abstract}
On October 6, 2018, the US Senate confirmed Brett Kavanaugh with the narrowest margin for a successful confirmation since 1881 and where the senators voted overwhelmingly along party lines.  In this paper, we examine whether the political polarization in the Senate is reflected among the general public.  To do so, we analyze the views of more than 128 thousand Twitter users.  We show that users supporting or opposing Kavanaugh's nomination were generally using divergent hashtags, retweeting different Twitter accounts, and sharing links from different websites.  We also examine characterestics of both groups.
\end{abstract}

\section{Introduction}
On October 6, 2018, the US senate confirmed Brett Kavanaugh to become a justice on the US Supreme Court with a 50 to 48 vote that was mostly along party lines.  This was the closest successful confirmation since the Stanley Matthews confirmation in 1881\footnote{\url{https://www.senate.gov/pagelayout/reference/nominations/Nominations.htm}}.  In this paper, we examine whether the political polarization at play in the US Senate between Republicans, who overwhelmingly voted for Kavanaugh, and Democrats, who overwhelmingly voted against him, reflects polarization among the general public.  We analyze more than 23 million tweets related to the Kavanaugh confirmation.  Initially, we semi-automatically tag more than 128 thousand Twitter users as supporting or opposing his confirmation.  Next, we bucket hashtags, retweeted accounts, and cited websites according to how strongly they are associated with those who support or oppose.  Further, we visualize users according to their similarity based on their hashtag usage, retweeted accounts, and cited websites.  All our analysis show strong polarization between both camps. Lastly, we highlight some of the main differences between both groups.

\section{Timeline}
On July 9, 2018, Brett Kavanaugh (BK), a US federal judge, was nominated by the US president Donald Trump to serve as a justice on the US supreme court to replace outgoing Justice Anthony Kennedy\footnote{\url{https://en.wikipedia.org/wiki/Brett_Kavanaugh}}. His nomination was marred by controversy with Democrats complaining that the White House withheld documents pertaining to BK's record and later a few women including a University of California professor accused him of sexual assault. The accusations of sexual misconducted led to a public congressional hearing on September 27, 2018 and a subsequent investigation by the Federal Bureau of Investigation (FBI).  The US Senate voted to confirm BK to a seat on the Supreme Court on October 6 with a 50--48 vote, which mostly aligned with party loyalties.  BK was sworn in later the same day.

\section{Data Collection}
We collected tweets pertaining to the nomination of BK in two different time epochs, namely September 28-30, which were the three days following the congressional hearing concerning the sexual assault allegation against BK, and October 6-9, which included the day the Senate voted to confirm BK and the following three days.  We collected tweets using the twarc toolkit\footnote{\url{https://github.com/edsu/twarc}}, where we used both the search and filtering interfaces to find tweets containing any of the following keywords: Kavanaugh, Ford, Supreme, judiciary, Blasey, Grassley, Hatch, Graham, Cornyn, Lee, Cruz, Sasse, Flake, Crapo, Tillis, Kennedy, Feinstein, Leahy, Durbin, Whitehouse, Klobuchar, Coons, Blumenthal, Hirono, Booker, or Harris.  The keywords include the BK's name, his main accuser, and the names of the members of the Senate's Judiciary Committee.  The per day breakdown of the collected tweets is as follows:
\begin{center}
\begin{tabular}{c|r}
Date     &   Count   \\ \hline
28-Sep	&	5,961,549	\\
29-Sep	&	4,815,160	\\
30-Sep	&	1,590,522	\\ \hline
subtotal	&	12,367,231	\\ \hline \hline
6-Oct	&	2,952,581	\\
7-Oct	&	3,448,315	\\
8-Oct	&	2,761,036	\\
9-Oct	&	1,687,433	\\ \hline
subtotal	&	10,849,365	\\ \hline \hline
Total	&	23,216,596	\\
\end{tabular}
\end{center}

In the process we collected 23 million tweets that were authored by 687,194 users.  Our first step was to label as many users as possible by their stance as supporting (SUPP) or opposing (OPP) the confirmation of BK.  The labeling process was done in three steps, namely:
\begin{itemize}
\item \textbf{Manual labeling of users.} We manually labeled 43 users who had the most number of tweets in our collection. Of them, the SUPP users were 29 compared to 12 OPP users.  The two remaining users were either neutral or spammers.  

\item \textbf{Label propagation.} Label propagation automatically labels users based on their retweet behavior \cite{darwish2017predicting,kutlu2018devam,magdy2016isisisnotislam}.  The intuition behind this method is that users that retweet the same tweets on a topic most likely share the same stance. Given that many of the tweets in our collection were actually retweets or duplicates of other tweets, we labeled users who retweeted 15 or more tweets that were authored or retweeted by the SUPP group or 7 or more times by OPP group and no retweets from the other side as SUPP or OPP respectively.  We elected to increase the minimum number for the SUPP group as they were over represented in the initial manually labeled set.  We iteratively performed such label propagation 4 times, which is when label propagation stopped labeling new accounts.  After the last iteration, we were able to label 65,917 users of which 26,812 were SUPP and 39,105 were OPP. Since we don't have golden labels to compare against, we opted to spot check the results. Thus, we randomly selected 10 automatically labeled accounts, and all of them were labeled correctly. This is intended as a sanity check.  A larger random sample is required for a more thorough evaluation.  Further, this labeling methodology naturally favors users who actively discuss a topic and who are likely to hold strong views. 

\item \textbf{Retweet based-classification.} We used the labeled users to train a classification model to guess the stances of users who retweeted at least 20 different accounts, which were users who were actively tweeting about the topic.  For classification, we used the FastText classification toolkit, which is an efficient deep neural network classifier that has been shown to be effective for text classification \cite{joulin2016bag}.  We used the accounts that each user retweeted as features.  Strictly using the retweeted accounts has been shown to be effective for stance classification \cite{magdy2016isisisnotislam}.  To keep precision high, we only trusted the classification of users where the classifier was more than 90\% confident.  Doing so, we increased the number of labeled users to 128,096, where 57,118 belonged to the SUPP group with 13,095,422 tweets and 70,978 belonged to the OPP group with 12,510,134 tweets.  Again, we manually inspected 10 random users who were automatically tagged and all of them were classified correctly. It is noteworthy that the relative number of SUPP to OPP users in not necessarily meaningful.  To determine the exact ratio, we would need to determine the stances of a large sample of users.

\end{itemize}

\section{Data Analysis}
Next we analyzed the data to ascertain the differences in interests and focus between both groups as expressed using three elements, namely the hashtags that they use, the accounts they retweet, and the websites that they cite (share content from). Doing so can provide valuable insights into both groups \cite{darwish2017predicting,darwish2017trump}.  For all three elements, we bucketed them into five bins reflecting how strongly they are associated with the SUPP and OPP groups.  These bins are: strong SUPP, SUPP, Neutral, OPP, and strong OPP.  To perform the bucketing, we used the so-called valence score \cite{conover2011political}.  The valence score for an element $e$ is computed as follows:

\begin{equation}
    V(e) = 2 \frac{
    \frac{tf_{SUPP}}{total_{SUPP}}}
    {\frac{tf_{SUPP}}{total_{SUPP}} + \frac{tf_{OPP}}{total_{OPP}}} -1
    \label{eq:valence}
\end{equation}
where $tf$ is the \textit{frequency} of the element in either the SUPP or OPP tweets and $total$ is the sum of all $tf$s for either the SUPP or OPP tweets. We accounted for all elements that appeared in at least 100 tweets.  Since the value of valence varies between -1 (strong OPP) to +1 (strong SUPP), we divided the ranged into 5 equal bins: strong OPP [-1.0 -- -0.6), OPP [-0.6 -- -0.2), Neutral [-0.2 -- 0.2), SUPP [0.2 -- 0.6), and strong SUPP [0.6 -- 1.0).

\begin{figure*}
    \centering
    \includegraphics[width=0.6\linewidth]{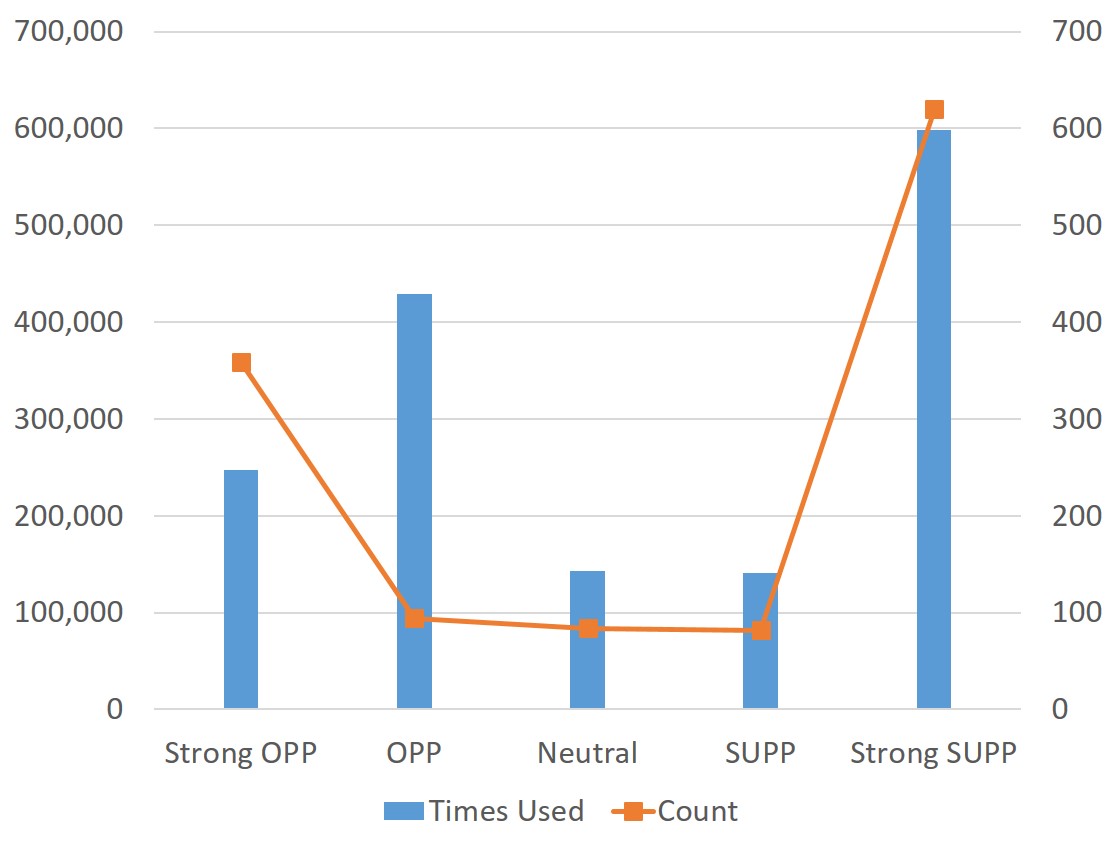}
    \caption{Count of hashtags and the number of times they are used for different valence bins}
    \label{fig:hashtagValence}
\end{figure*}

\begin{figure*}
    \centering
    \includegraphics[width=0.6\linewidth]{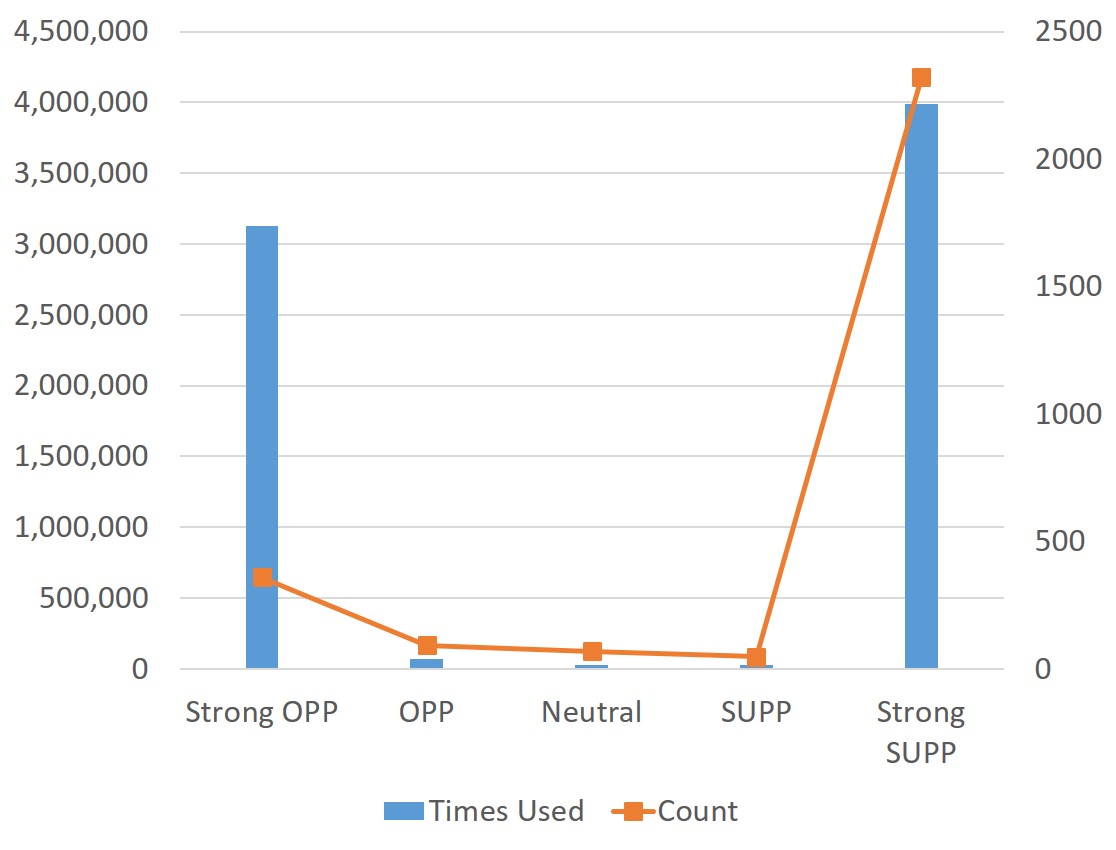}
    \caption{Count of retweeted account and the number of times they are retweeted for different valence bins}
    \label{fig:RTValence}
\end{figure*}

\begin{figure*}
    \centering
    \includegraphics[width=0.6\linewidth]{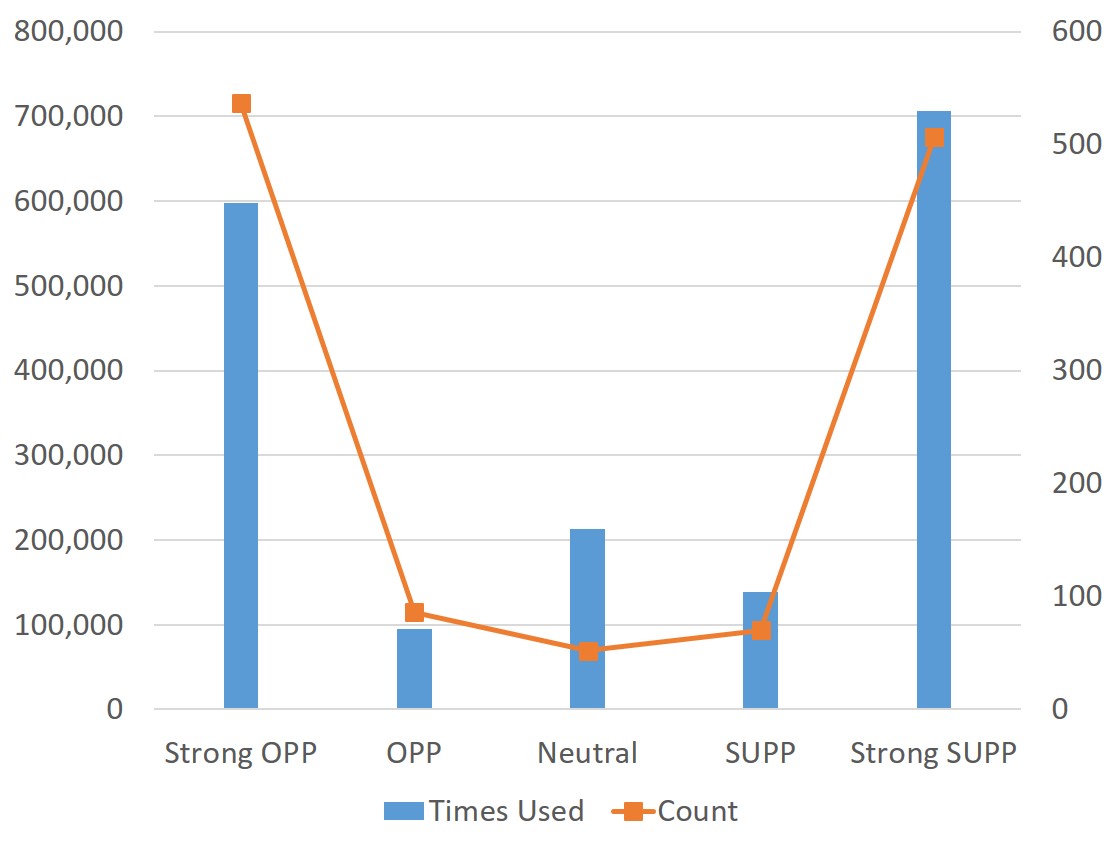}
    \caption{Count of websites and the number of times they are cited for different valence bins}
    \label{fig:URLValence}
\end{figure*}

Figures \ref{fig:hashtagValence}, \ref{fig:RTValence}, and \ref{fig:URLValence} respectively provide the number of different hashtags, retweeted accounts, and cited websites that appear for all five bins along with the number of tweets in which they are used.  As the figures show, there is strong polarization between both camps.  Polarization is most evident in the accounts that they retweet and the websites that they share content from, where ``strong SUPP'' and ``strong OPP'' groups dominate in terms of the number of elements and their frequency.

Tables \ref{table:topHashTags}, \ref{table:topRT}, and \ref{table:topURL} respectively show the 15 most commonly used hashtags, retweeted accounts, and most cited websites for each of the valence bins.  Since the ``Strong SUPP'' and ``strong OPP'' groups are most dominant, we focus here on their main characteristics. 

\begin{table*}
\begin{center}
\begin{scriptsize}
    \begin{tabular}{c|c|c|c|c}
         Strong SUPP	&	SUPP	&	Neutral	&	OPP	&	Strong OPP	\\ \hline
MAGA	&	SCOTUS	&	KavanaughHearings	&	Kavanaugh	&	DelayTheVote	\\
Winning	&	ChristineBlaseyFord	&	KavanaughVote	&	MeToo	&	StopKavanaugh	\\
ConfirmKavanaugh	&	kavanaughconfirmation	&	Breaking	&	BrettKavanaugh	&	GOP	\\
ConfirmKavanaughNow	&	Ford	&	FBI	&	Trump	&	BelieveSurvivors	\\
walkaway	&	kavanaughconfirmed	&	flake	&	republican	&	SNLPremiere	\\
JusticeKavanaugh	&	SaturdayMorning	&	JeffFlake	&	DrFord	&	SNL	\\
QAnon	&	TedCruz	&	SupremeCourt	&	KavanaghHearing	&	TheResistance	\\
Democrats	&	FridayFeeling	&	Grassley	&	BelieveWomen	&	Resist	\\
TXSen	&	HimToo	&	LindseyGraham	&	Republicans	&	Voteno	\\
Midterms	&	TCOT	&	KavanaughHearing	&	RT	&	SusanCollins	\\
LDTPoll	&	SundayMorning	&	WhiteHouse	&	NeverTrump	&	Vote	\\
FoxNews	&	USA	&	America	&	Resistance	&	SmartNews	\\
WWG1WGA	&	KavanaughConfirmationHearing	&	WomensMarch	&	Kavanagh	&	JulieSwetnick	\\
Feinstein	&	RememberInNovember	&	Senate	&	BrettKavanuagh	&	KavaNo	\\
VoteRed2018	&	KavanaughFord	&	FBIInvestigation	&	Collins	&	VoteBlue	\\
    \end{tabular}
    \caption{Top hashtags}
    \label{table:topHashTags}
    \end{scriptsize}
\end{center}
\end{table*}

\begin{table*}
\begin{center}
\begin{scriptsize}
    \begin{tabular}{c|c|c|c|c}
         Strong SUPP	&	SUPP	&	Neutral	&	OPP	&	Strong OPP	\\ \hline
realDonaldTrump	&	PollingAmerica	&	RiegerReport	&	AP	&	krassenstein	\\
mitchellvii	&	cspan	&	lachlan	&	CBSNews	&	kylegriffin1	\\
dbongino	&	JenniferJJacobs	&	Sen\_JoeManchin	&	Reuters	&	KamalaHarris	\\
charliekirk11	&	JerryDunleavy	&	AaronBlake	&	USATODAY	&	SenFeinstein	\\
FoxNews	&	JulianSvendsen	&	WSJ	&	Phil\_Mattingly	&	EdKrassen	\\
RealJack	&	jamiedupree	&	markknoller	&	dangercart	&	thehill	\\
DineshDSouza	&	CNNSotu	&	Bencjacobs	&	WalshFreedom	&	MichaelAvenatti	\\
Thomas1774Paine	&	AlBoeNEWS	&	lawrencehurley	&	4YrsToday	&	SethAbramson	\\
AnnCoulter	&	elainaplott	&	AureUnnie	&	byrdinator	&	funder	\\
foxandfriends	&	AlanDersh	&	choi\_bts2	&	MittRomney	&	Lawrence	\\
JackPosobiec	&	FoxNewsResearch	&	happinesspjm	&	MacFarlaneNews	&	tedlieu	\\
paulsperry\_	&	SCOTUSblog	&	threadreaderapp	&	TexasTribune	&	MSNBC	\\
RealCandaceO	&	W7VOA	&	soompi	&	HotlineJosh	&	JoyceWhiteVance	\\
McAllisterDen	&	scotusreporter	&	ya\_\_mi\_\_\_ya\_mi	&	BBCWorld	&	tribelaw	\\
IngrahamAngle	&	CraigCaplan	&	savtwopointoh	&	Mediaite	&	Amy\_Siskind	\\
    \end{tabular}
        \caption{Top retweeted accounts}
    \label{table:topRT}
    \end{scriptsize}
\end{center}
\end{table*}

\begin{table*}
\begin{center}
\begin{scriptsize}
    \begin{tabular}{c|c|c|c|c}
         Strong SUPP	&	SUPP	&	Neutral	&	OPP	&	Strong OPP	\\ \hline
thegatewaypundit.com	&	usatoday.com	&	dr.ford	&	nytimes.com	&	hill.cm	\\
foxnews.com	&	mediaite.com	&	twitter.com/michaelavenatti/	&	twitter.com/thehill/	&	wapo.st	\\
fxn.ws	&	twitter.com/realdonaldtrump/	&	dailym.ai	&	thehill.com	&	washingtonpost.com	\\
dailycaller.com	&	theweek.com	&	lawandcrime.com	&	politi.co	&	rawstory.com	\\
breitbart.com	&	twitter.com/lindseygrahamsc/	&	nypost.com	&	abcn.ws	&	vox.com	\\
twitter.com/foxnews/	&	nyp.st	&	twitter.com/donaldjtrumpjr/	&	usat.ly	&	huffingtonpost.com	\\
thefederalist.com	&	twitter.com/senfeinstein/	&	twitter.com/senjudiciary/	&	axios.com	&	nyti.ms	\\
westernjournal.com	&	twitter.com/kamalaharris/	&	twitter.com/mediaite/	&	politico.com	&	nbcnews.com	\\
twitter.com/politico/	&	twitter.com/newsweek/	&	c-span.org	&	reut.rs	&	cnn.com	\\
ilovemyfreedom.org	&	twitter.com/natesilver538/	&	twitter.com/gop/	&	po.st	&	apple.news	\\
chicksonright.com	&	twitter.com/mcallisterden/	&	twitter.com/kendilaniannbc/	&	twitter.com/nbcnews/	&	dailykos.com	\\
hannity.com	&	twitter.com/foxandfriends/	&	realclearpolitics.com	&	twitter.com/brithume/	&	cnn.it	\\
nationalreview.com	&	chn.ge	&	twitter.com/senatorcollins/	&	dailymail.co.uk	&	palmerreport.com	\\
dailywire.com	&	fastcompany.com	&	twitter.com/samstein/	&	twitter.com/cnnpolitics/	&	hillreporter.com	\\
bigleaguepolitics.com	&	rollcall.com	&	twitter.com/johncornyn/	&	abcnews.go.com	&	newyorker.com	\\

    \end{tabular}
        \caption{Top websites}
    \label{table:topURL}
    \end{scriptsize}
\end{center}
\end{table*}

For the \textbf{``Strong SUPP''} group, the hashtags can be split into the following topics (in order of importance as determined by frequency):
\begin{itemize}
    \item \textbf{Trump related:} \#MAGA (Make America Great Again), \#Winning.
    \item \textbf{Pro Kavanaugh confirmation:} \#ConfirmKavanaugh, \#ConfirmKavanaughNow, \#JusticeKavanaugh.
    \item \textbf{Anti-Democratic Party:} \#walkAway (campaign to walkaway from liberalism), \#Democrats, \#Feinstein
    \item \textbf{Conspiracy theories:} \#QAnon (an alleged Trump administration leaker), \#WWG1WGA (Where We Go One We Go All)
    \item \textbf{Midterm elections:} \#TXSen (Texas senator Ted Cruz), \#Midterms, \#VoteRed2018 (vote Republican)
    \item \textbf{Conservative media:} \#FoxNews, \#LDTPoll (Lou Dobbs (Fox News) on Twitter poll)
\end{itemize}
It is interesting to see hashtags expressing support for Trump (\#MAGA and \#Wining) feature more prominently than those that indicate support for BK. Further, anti-Democratic party hashtags (ex. \#WalkAway and \#Democrats) and hashtags related to conspiracy theories (ex. \#QAnon) may also indicate polarization.

Retweeted accounts reflect a similar trend to that of hashtags.  The retweeted accounts can be grouped as follows (in order of importance):
\begin{itemize}
    \item \textbf{Trump related:} \@realDonaldTrump, \@mitchellvii (Bill Mitchell -- social media personality who staunchly supports Trump), \@RealJack (Jack Murphy -- co-owner of \url{ILoveMyFreedom.org} (pro-Trump website)), \@DineshDSouza (Dinesh D'Souza -- commentator and film maker)
    \item \textbf{Conservative media:} \@dbongino (Dan Bongino -- author with podcast), \@FoxNews, \@FoxAndFriends (Fox News), \@JackPosobiec (Jack Posobiec -- One America News Network), \@IngrahamAngle (Laura Ingraham -- Fox News)
    \item \textbf{Conservative/Republican personalities:} \@charliekirk11 (Charlie Kirk -- founder of Turning Point USA), \@AnnCoulter (Ann Coulter -- author and commentator), \@Thomas1774Paine (Thomas Paine -- author), \@paulsperry\_ (Paul Sperry -- author and media personality), \@RealCandaceO (Candace Owens -- Turning Point USA), \@McAllisterDen (D C McAllister -- commentator)
\end{itemize}
The list above show that specifically pro-Trump accounts featured even more prominently than conservative accounts. 

For the same group, cited website were generally right-leaning, with some of them being far-right and most of them having mixed credibility.  We list here their leanings and their credibility levels according to the Media Bias/Fact Check website\footnote{\url{mediabiasfactcheck.com}}:
\begin{tabular}{l|c|c}
Source & Bias & Credibility \\ \hline
thegatewaypundit.com & far right & questionable \\
foxnews.com & right & mixed \\
fxn.ws (Fox News) & right & mixed \\
dailycaller.com & right & mixed \\
breitbart.com & far right & questionable \\
twitter.com/foxnews/ & right & mixed \\
thefederalist.com & right & high \\
westernjournal.com & right & mixed \\
twitter.com/politico/ & left-center & high \\
ilovemyfreedom.org & far right & questionable \\
chicksonright.com & not listed & not listed\\
hannity.com (Fox News) & not listed & not listed \\
nationalreview.com & right & mixed \\
dailywire.com & right & mixed \\
bigleaguepolitics.com & right & mixed \\ \hline
\end{tabular}
\\

For the \textbf{``strong OPP''} group, the top hashtags can be topically grouped as follows (in order of importance):
\begin{itemize}
    \item \textbf{Anti Kavanaugh:} \#DelayTheVote, \#StopKavanaugh, \#KavaNo (no to Kavanaugh), \#voteNo.
    \item \textbf{Republican Party related:} \#GOP, \#SusanCollins (GOP senator voting for Kavanaugh).
    \item \textbf{Sexual assault related:} \#BelieveSurvivors, \#JulieSwetnick (woman accusing Kavanaugh).
    \item \textbf{Media related:} \#SNLPremiere (Saturday Night Live satirical show), \#SNL, \#SmartNews (anti-Trump/GOP news)
    \item \textbf{Anti Trump:} \#TheResistance, \#Resist
    \item \textbf{Midterms:} \#vote, \#voteBlue (vote democratic)
\end{itemize}
As the list shows that the most prominent hashtags were related to opposition to the confirmation of BK.  Opposition to the Republican Party (\#GOP) and Trump (\#TheResistance) may indicate polarization.

As for their retweeted accounts, media related accounts dominated the list.  The remaining accounts belonged to prominent Democratic Party officials and anti-Trump accounts.  The details are as follows (in order of importance): 
\begin{itemize}
    \item \textbf{Media related:} \@krassenstein (Brian Krassenstein -- \url{HillReporter.com}), \@kylegriffin1 (Kyle Griffin -- MSNBC producer), \@EdKrassen (Ed Krassenstein -- \url{HillReporter.com}), \@theHill, \@funder (Scott Dworkin -- Dworkin Report and Democratic Coallition), \@Lawrence (Lawrence O'Donnell -- MSNBC), \@MSNBC, \@JoyceWhiteVance (Joyce Alene -- law professor and MSNBC contributor), \@Amy\_Siskind (Amy Siskind -- The Weekly List)
    \item \textbf{Democratic Party:} \@KamalaHarris (Senator Kamala Harris), \@SenFeinstein (Senator Dianne Feinstein), \@tedlieu (Representative Ted Lieu)
    \item \textbf{Anti Kavanaugh:} \@MichaelAvenatti (Michael Avenatti -- lawyer of Kavanaugh accuser)
    \item \textbf{Anti Trump:} \@SethAbramson (Seth Abramson -- author of Proof of Collusion), \@tribelaw (Laurence Tribe -- Harvard Professor and author of ``To End a Presidency'')
\end{itemize}

Concerning cited websites, they mostly left or left-of-center leaning sources.  The credibility of the sources were generally higher than those for the ``strong SUPP'' group.  Their details are as follows:
\begin{tabular}{l|c|c}
Source & Bias & Credibility \\ \hline
hill.cm (the Hill) & left-center & high \\
wapo.st (Washington Post) & left-center & high \\
washingtonpost.com & left-center & high \\
rawstory.com & left & mixed \\
vox.com & left & high \\
huffingtonpost.com & left & high \\
nyti.ms (New York Times) & left-center & high\\
nbcnews.com & left-center & high\\
cnn.com & left & mixed \\
apple.news & not list & not listed\\
dailykos.com & left & mixed \\
cnn.it & left & mixed \\
palmerreport.com & left & mixed\\
hillreporter.com & not listed & not listed\\
newyorker.com & left & high\\ \hline
\end{tabular}
\\

Lastly, we examined the retweeted accounts and cited websites that appeared in the ``neutral'' bin, meaning that they were shared by both SUPP and OPP groups.  We were interested specifically in the media accounts.  They are as follows:
\begin{itemize}
    \item \textbf{Retweeted accounts:} \@RiegerReport (JM Rieger -- Washington Post video editor), \@lachlan (Lachlan Markay -- Daily Beast reporter), \@AaronBlake (Aaron Blake -- Washington Post reporter), \@WSJ (Wall Street Journal), \@markknoller (Mark Knoller -- CBS News correspondent), \@Bencjacobs (Ben Jacobs -- the Guardian), \@lawrencehurley (Lawrence Hurley -- Reuters reporter) \item \textbf{Websites/URLs:} dailym.ai (Daily Mail), LawAndCrime.com (Live trial network), nypost.com (New York Post), twitter.com/mediaite (MediaITE), c-span.org (CSPAN), twitter.com/kendilaniannbc/ (Ken Dilanian -- NBC News), RealClearPolitics.com, twitter.com/samstein/ (Sam Stein -- Daily Beast/MSNBC newsletter)
\end{itemize}
One interesting thing in this list is that although some sources such as the Washington Post was most cited by the ``strong OPP'' group, some Post's reporters featured prominently for the ``neutral'' group.

Three politicians also appeared in the neutral column, namely Sen\_JoeManchin (Democratic Senator Joe Manchin from West Virginia), SenatorCollins (Republican Senator Susan Collins from Maine), and JohnCornyn (Republican Senator John Cornyn from Texas). Incidentally, all three of them voted to confirm Kavanaugh.

Next, we examined polarization in terms of the similarity between users based on the three aforementioned elements, hashtags, retweeted accounted, and cited websites.  We computed the cosine similarity between users based on their usage of the different elements.  For user vectors given element types, we normalized the occurrence values of elements to ensure that sum of values in the vector add up to 1.  For example, if user ``A'' uses three hashtags with frequencies 5, 100, and 895, the corresponding feature values would be 5/1,000, 100/1,000, and 895/1,000, where 1,000 is the sum of the frequencies.  We computed the similarities of users who used a minimum of 10 elements to ensure a minimum level of engagement in the topic.  Next, we visualized the network of users using the NetworkX toolkit\footnote{\url{https://networkx.github.io/}} which uses Fruchterman-Reingold force-directed algorithm to space out the nodes. In essence, the distance between users would correlate positively with their similarity \cite{darwish2017seminar}.  

\begin{figure*}
    \centering
    \includegraphics[width=0.55\linewidth]{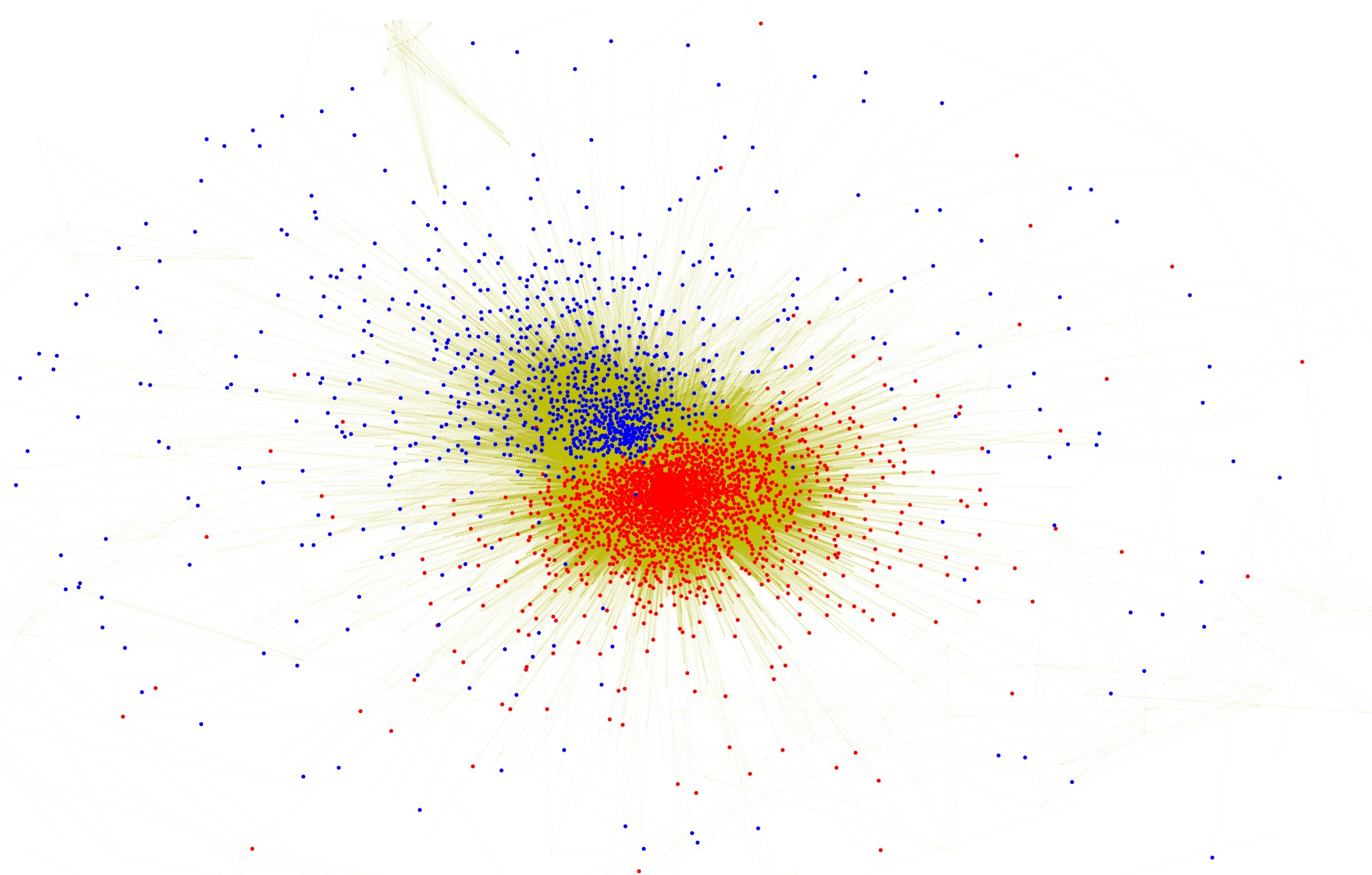}
    \caption{User similarity given the hashtags they used.  Red is for SUPP users and Blue is for OPP users.}
    \label{fig:hashtagForce}
\end{figure*}

\begin{figure*}
    \centering
    \includegraphics[width=0.55\linewidth]{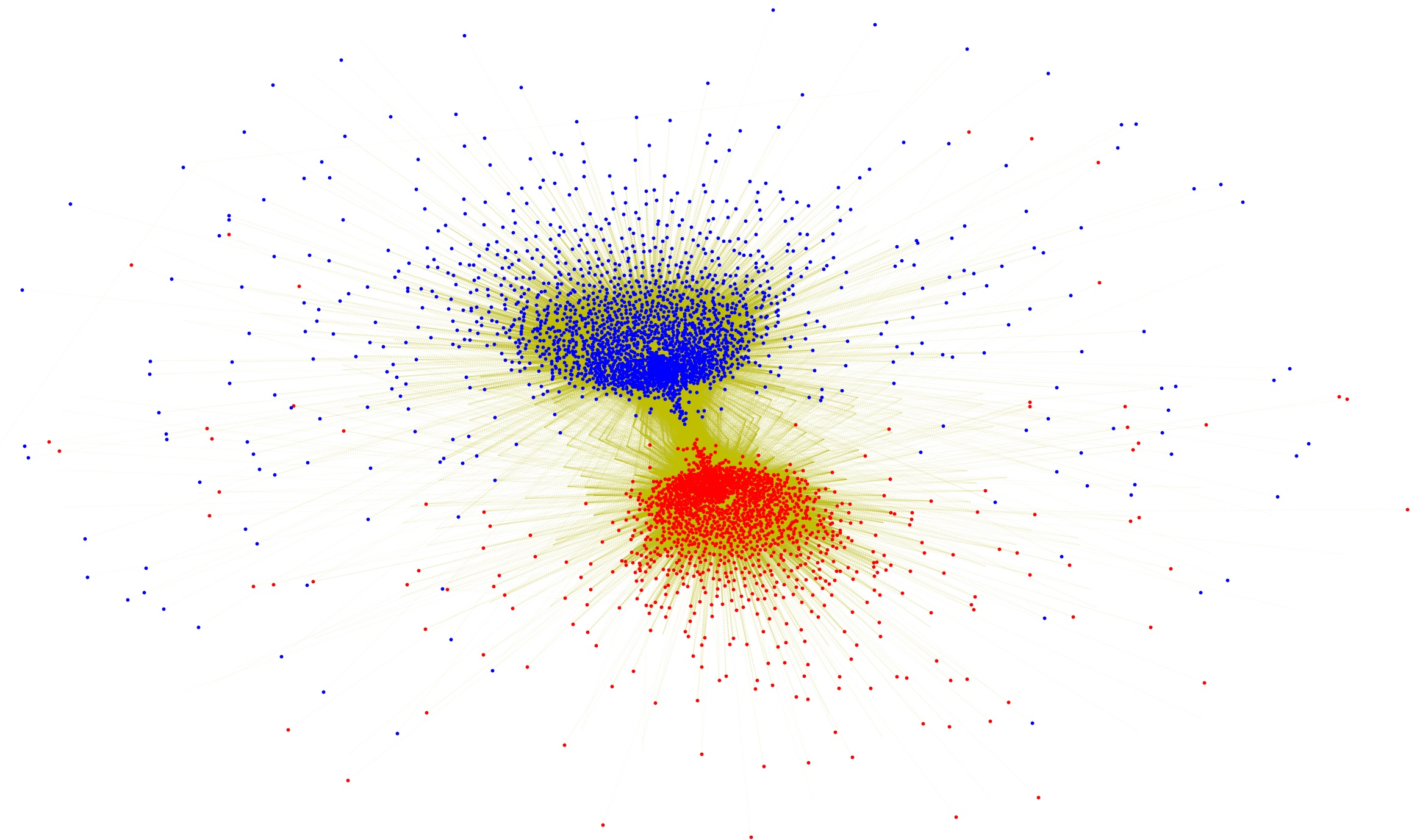}
    \caption{User similarity given the accounts they retweeted.  Red is for SUPP users and Blue is for OPP users.}
    \label{fig:RTForce}
\end{figure*}

\begin{figure*}
    \centering
    \includegraphics[width=0.55\linewidth]{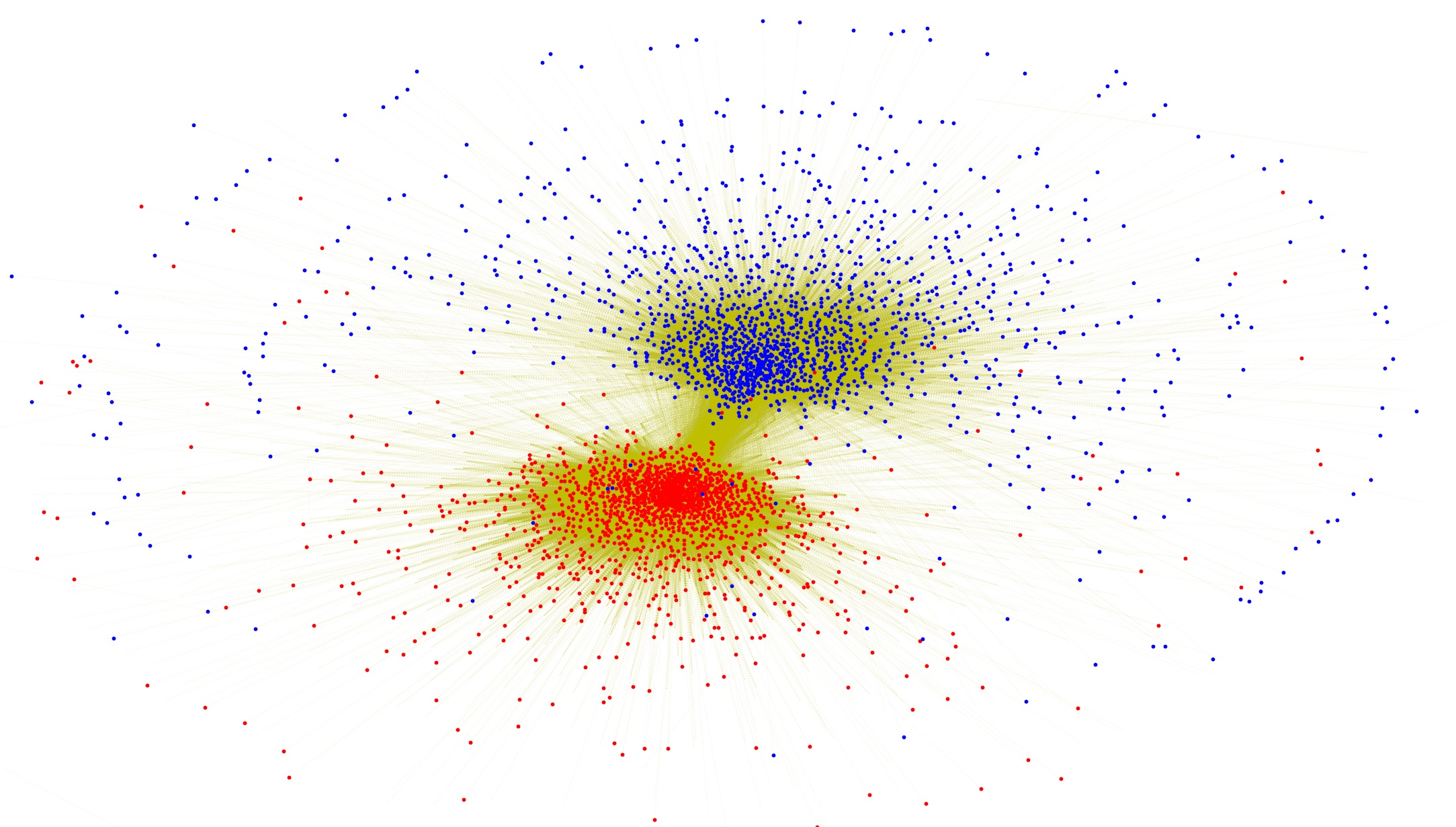}
    \caption{User similarity given the websites they cited.  Red is for SUPP users and Blue is for OPP users.}
    \label{fig:URLForce}
\end{figure*}

Figures \ref{fig:hashtagForce}, \ref{fig:RTForce}, and \ref{fig:URLForce} respectively show the similarity between 5,000 randomly selected users using hashtags, retweeted accounts, and cited websites.  The red dots represent SUPP users and the blue dots represent OPP users.  As the Figures clearly show, both groups are clearly separable, which indicates polarization. For retweeted accounts and cited websites, polarization is more evident.

\section{Conclusion}

In this paper, we examined whether the political polarization between Republican and Democratic senators on display during the Supreme Court confirmation hearings of judge Brett Kavanaugh reflects polarization of social media users.  To do so, we analyzed more than 128 thousand Twitter users.  In the process, we showed that those who support and oppose the confirmation of Kavanaugh were generally using divergent hashtags and were following different Twitter accounts and websites. 

\bibliographystyle{aaai}
\bibliography{references}

\end{document}